\documentclass[12pt,a4]{article}

\begin{document}
\thispagestyle{empty}
\vspace*{-1.5cm}
\hfill {\small KL--TH 98/9} \\[8mm]

\message{reelletc.tex (Version 1.0): Befehle zur Darstellung |R  |N, Aufruf z.B. \string\bbbr}
%
%
\message{reelletc.tex (Version 1.0): Befehle zur Darstellung |R  |N, Aufruf z.B. \string\bbbr}
%
%
%
%
%
\font \smallescriptscriptfont = cmr5
\font \smallescriptfont       = cmr5 at 7pt
\font \smalletextfont         = cmr5 at 10pt
\font \tensans                = cmss10
\font \fivesans               = cmss10 at 5pt
\font \sixsans                = cmss10 at 6pt
\font \sevensans              = cmss10 at 7pt
\font \ninesans               = cmss10 at 9pt
\newfam\sansfam
\textfont\sansfam=\tensans\scriptfont\sansfam=\sevensans
\scriptscriptfont\sansfam=\fivesans
\def\sans{\fam\sansfam\tensans}
\def\bbbr{{\rm I\!R}} 
\def\bbbn{{\rm I\!N}} 
\def\bbbE{{\rm I\!E}} 
\def\bbbm{{\rm I\!M}}
\def\bbbh{{\rm I\!H}}
\def\bbbk{{\rm I\!K}}
\def\bbbd{{\rm I\!D}}
\def\bbbp{{\rm I\!P}}
\def\bbbone{{\mathchoice {\rm 1\mskip-4mu l} {\rm 1\mskip-4mu l}
{\rm 1\mskip-4.5mu l} {\rm 1\mskip-5mu l}}}
\def\bbbc{{\mathchoice {\setbox0=\hbox{$\displaystyle\rm C$}\hbox{\hbox
to0pt{\kern0.4\wd0\vrule height0.9\ht0\hss}\box0}}
{\setbox0=\hbox{$\textstyle\rm C$}\hbox{\hbox
to0pt{\kern0.4\wd0\vrule height0.9\ht0\hss}\box0}}
{\setbox0=\hbox{$\scriptstyle\rm C$}\hbox{\hbox
to0pt{\kern0.4\wd0\vrule height0.9\ht0\hss}\box0}}
{\setbox0=\hbox{$\scriptscriptstyle\rm C$}\hbox{\hbox
to0pt{\kern0.4\wd0\vrule height0.9\ht0\hss}\box0}}}}

\def\bbbe{{\mathchoice {\setbox0=\hbox{\smalletextfont e}\hbox{\raise
0.1\ht0\hbox to0pt{\kern0.4\wd0\vrule width0.3pt height0.7\ht0\hss}\box0}}
{\setbox0=\hbox{\smalletextfont e}\hbox{\raise
0.1\ht0\hbox to0pt{\kern0.4\wd0\vrule width0.3pt height0.7\ht0\hss}\box0}}
{\setbox0=\hbox{\smallescriptfont e}\hbox{\raise
0.1\ht0\hbox to0pt{\kern0.5\wd0\vrule width0.2pt height0.7\ht0\hss}\box0}}
{\setbox0=\hbox{\smallescriptscriptfont e}\hbox{\raise
0.1\ht0\hbox to0pt{\kern0.4\wd0\vrule width0.2pt height0.7\ht0\hss}\box0}}}}

\def\bbbq{{\mathchoice {\setbox0=\hbox{$\displaystyle\rm Q$}\hbox{\raise
0.15\ht0\hbox to0pt{\kern0.4\wd0\vrule height0.8\ht0\hss}\box0}}
{\setbox0=\hbox{$\textstyle\rm Q$}\hbox{\raise
0.15\ht0\hbox to0pt{\kern0.4\wd0\vrule height0.8\ht0\hss}\box0}}
{\setbox0=\hbox{$\scriptstyle\rm Q$}\hbox{\raise
0.15\ht0\hbox to0pt{\kern0.4\wd0\vrule height0.7\ht0\hss}\box0}}
{\setbox0=\hbox{$\scriptscriptstyle\rm Q$}\hbox{\raise
0.15\ht0\hbox to0pt{\kern0.4\wd0\vrule height0.7\ht0\hss}\box0}}}}

\def\bbbt{{\mathchoice {\setbox0=\hbox{$\displaystyle\rm
T$}\hbox{\hbox to0pt{\kern0.3\wd0\vrule height0.9\ht0\hss}\box0}}
{\setbox0=\hbox{$\textstyle\rm T$}\hbox{\hbox
to0pt{\kern0.3\wd0\vrule height0.9\ht0\hss}\box0}}
{\setbox0=\hbox{$\scriptstyle\rm T$}\hbox{\hbox
to0pt{\kern0.3\wd0\vrule height0.9\ht0\hss}\box0}}
{\setbox0=\hbox{$\scriptscriptstyle\rm T$}\hbox{\hbox
to0pt{\kern0.3\wd0\vrule height0.9\ht0\hss}\box0}}}}

\def\bbbs{{\mathchoice
{\setbox0=\hbox{$\displaystyle     \rm S$}\hbox{\raise0.5\ht0\hbox
to0pt{\kern0.35\wd0\vrule height0.45\ht0\hss}\hbox
to0pt{\kern0.55\wd0\vrule height0.5\ht0\hss}\box0}}
{\setbox0=\hbox{$\textstyle        \rm S$}\hbox{\raise0.5\ht0\hbox
to0pt{\kern0.35\wd0\vrule height0.45\ht0\hss}\hbox
to0pt{\kern0.55\wd0\vrule height0.5\ht0\hss}\box0}}
{\setbox0=\hbox{$\scriptstyle      \rm S$}\hbox{\raise0.5\ht0\hbox
to0pt{\kern0.35\wd0\vrule height0.45\ht0\hss}\raise0.05\ht0\hbox
to0pt{\kern0.5\wd0\vrule height0.45\ht0\hss}\box0}}
{\setbox0=\hbox{$\scriptscriptstyle\rm S$}\hbox{\raise0.5\ht0\hbox
to0pt{\kern0.4\wd0\vrule height0.45\ht0\hss}\raise0.05\ht0\hbox
to0pt{\kern0.55\wd0\vrule height0.45\ht0\hss}\box0}}}}

\def\bbbz{{\mathchoice {\hbox{$\sans\textstyle Z\kern-0.4em Z$}}
{\hbox{$\sans\textstyle Z\kern-0.4em Z$}}
{\hbox{$\sans\scriptstyle Z\kern-0.3em Z$}}
{\hbox{$\sans\scriptscriptstyle Z\kern-0.2em Z$}}}}
\setlength{\topmargin}{-1.5cm}
\setlength{\textheight}{22cm}
\begin{center}
{\large\bf Construction of exactly solvable quantum models of  \\
Calogero and Sutherland type with \\
translation invariant four-particle interactions}\\
\vspace{0.5cm}
{\large O. Haschke and W. R\"uhl}\\
Department of Physics, University of Kaiserslautern, P.O.Box 3049\\
67653 Kaiserslautern, Germany \\
\vspace{5cm}
\begin{abstract}
We construct exactly solvable models for four particles moving on a real line or 
on a circle with translation invariant two-- and four--particle interactions. The 
Sutherland type model seems to be hitherto unknown.
\end{abstract}
\vspace{3cm}
{\it July 1998}
\end{center}
\newpage

\section{Introduction}
All quantum mechanical models of the infinite sequences $A_n, B_n, C_n, BC_n, D_n$ and in 
addition some exceptional sequence models which were defined and proved to be completely 
integrable by Olshanetsky and Perelomov by the "Hamiltonian reduction method" \cite{1} 
possess limiting versions of Sutherland and Calogero type that can be shown to be exactly 
solvable by a simple trick. Namely, their Schr\"odinger operators can be transcribed into 
quadratic polynomials of related "hidden" Lie algebras. The spectrum and the eigenfunctions 
can then be determined from representation theory, e.g. using polynomial spaces .

This program was formulated in \cite{2} and successfully applied first to the $A_n$ sequence 
in \cite{3}. Then it was carried over to the other sequences and even the supersymmetric 
generalizations in \cite{4,5}.

Our aim is to turn the arguments around and to develop an algorithm which leads to new 
exactly solvable models. First investigations were presented in \cite{6} (hereafter 
quoted as (I)). The program contains two major and separate issues: to render a second 
order differential operator curvature free and to find a first order differential 
operator satisfying an integrability constraint. If both constraints are fulfilled, 
the whole operator can be transformed into a Schr\"odinger operator with a standard 
Laplacian as kinetic energy and a real potential as potential energy terms. 

In this work we apply this method to four-particle models. We start from the second order 
differential operator for $A_3$ known from \cite{3} (Calogero case) and \cite{6} (Sutherland case). 
After a change of the variable $\tau_3$ to $\lambda_3=\tau_3^2$ we test out all possible 
first order differential operators. This way we find a new solution leading to a four--particle 
potential which is automatically translation--invariant. In the Calogero case this model was known before \cite{7}, but the Sutherland case seems to be new.   

\section{The program}
We are interested here in the bound state spectrum of Schr\"odinger operators. The 
whole analysis is therefore performed in real spaces. Consider a flag of polynomial 
spaces $V_N, \, N \in \bbbz_{\ge}$
\begin{eqnarray}
& V_N = {\rm span} \left\{ x^{r_1}_1 x^{r_2}_2 ... x^{r_n}_n | r_1p_1 + r_2p_2 + ... + r_np_n \le N \right\} &\label{1} \\
&(p_i \in \bbbn)& \nonumber
\end{eqnarray}
We consider differential operators of first order
\begin{equation}
D^{(1)}_{[\vec{\alpha};a]} = x^{[\vec{\alpha}]} \frac{\partial}{\partial x_a}
\label{2}
\end{equation}
($\vec{\alpha}$ a multi-exponent) \\
and of second order
\begin{equation}
D^{(2)}_{[\vec{\alpha};a,b]} = x^{[\vec{\alpha}]} \frac{\partial^2}{\partial x_a \partial x_b}
\label{3}
\end{equation}
that leave each space $V_N$ invariant. If
\begin{equation}
\vec{p} = (1,1,...,1) 
\label{4}
\end{equation}
then the operators (\ref{2}) generate the full linear (inhomogeneous) group of $\bbbr_n$ 
and the operators of second order (\ref{3}) can be obtained as products from the first 
order operators, i.e. in (\ref{2})
\begin{equation}
\vec{\alpha} = e^{(a)}, \; e^{(a)}_b = \delta^a_b
\label{5}
\end{equation}
and in (\ref{3})
\begin{equation}
\vec{\alpha} = e^{(a)} + e^{(b)}
\label{6}
\end{equation}

Now we consider a candidate for a future Schr\"odinger operator
\begin{eqnarray}
D &=& - \sum_{\vec{\alpha},a,b} g_{[\vec{\alpha};a,b]} D^{(2)}_{[\vec{\alpha};a,b]} \nonumber \\
&+& \sum_{\vec{\beta},c} h_{[\vec{\beta};c]} D^{(1)}_{[\vec{\beta};c]}
\label{7}
\end{eqnarray}
The eigenvectors and values of $D$ in $V_N$ can be calculated easily by finite linear 
algebra methods (the number of eigenvector in $V_N$ may be smaller than $\dim V_N$, see (I)). 
If we want completely integrable models we must make sure that a complete set of involutive 
differential operators exists. For this task Lie algebraic methods may be very helpful.

The first step in transforming $D$ into a Schr\"odinger operator is to write it symmetrically
\begin{equation}
D = - \sum_{a,b} \frac{\partial}{\partial x_a} g^{-1}_{ab} (x) \frac{\partial}{\partial x_b} +
\sum_a r_a(x) \frac{\partial}{\partial x_a} 
\label{8}
\end{equation}
where
\begin{equation}
g^{-1}_{ab} = \sum_{\vec{\alpha}} g_{[\vec{\alpha};a,b]} x^{[\vec{\beta}]}
\label{9}
\end{equation}
We write $g^{-1}_{ab}$ because this is the inverse of a Riemann tensor. The Riemann tensor 
$g_{ab}$ is assumed to be curvature free. The task to make it so will not arise in this work. 
But we mention that we developed a minimal algorithm to solve this issue.

Following the notations of (I) we "gauge" the polynomial eigenfunctions $\varphi$ of $D$ by
\begin{equation}
\psi(x) = e^{-\chi(x)} \varphi(x)
\label{10}
\end{equation}
so that
\begin{equation}
e^{-\chi} D e^{+\chi} = - \frac{1}{\sqrt{g}} \sum_{a,b} \frac{\partial}{\partial x_a} 
(\sqrt{g} g^{-1}_{ab}) \frac{\partial}{\partial x_b} + W(x)
\label{11}
\end{equation}
$(g = (\det g^{-1})^{-1})$. \\
This is possible if and only if
\begin{equation}
 \sum_b g^{-1}_{ab} (x) \frac{\partial}{\partial x_b} [ 2 \chi - \ln \sqrt{g} ] = r_a(x)
\label{12}
\end{equation}
which implies integrability constraints on the functions $\{r_a(x)\}$. If they are fulfilled 
we obtain a "prepotential"
\begin{equation}
\rho = \ln P
\label{13}
\end{equation}
so that
\begin{equation}
\rho = 2 \chi - \ln \sqrt{g}
\label{14}
\end{equation}
In all cases studied, we found solutions for $\rho$ as follows. Let
\begin{equation}
\det g^{-1}(x) = \prod^r_{i=1} P_i(x)
\label{15}
\end{equation}
where $\{P_i(x)\}$ are different real polynomials. Then
\begin{equation}
\rho(x) = \sum^r_{i=1} \gamma_i \ln P_i(x)
\label{16}
\end{equation}
with free parameters $\gamma_i$ solves the requirement that $\{r_a(x)\}$ (\ref{12}) belong 
to differential operators leaving each $V_N$ invariant. In particular
\begin{equation}
r_a^{(i)}(x) = \frac{1}{P_i(x)} \sum_b g^{-1}_{ab}(x) \frac{\partial P_i}{\partial x_b}
\label{17}
\end{equation}
are polynomials, and each power $[\vec{\alpha}]$ is bounded by the requirement that the 
spaces $V_N(1)$ are invariant. Inserting (\ref{15}), (\ref{16}) 
in (\ref{14}) we obtain finally
 
\begin{equation}
\chi = \frac12 \sum^r_{i=1} (\gamma_i - \frac12) \ln P_i(x) 
\label{18}
\end{equation}
We will later see that in the case of the models of Calogero type a term
\begin{equation}
\gamma_0 \ln P_0(x)
\label{19}
\end{equation}
can be added to $\rho$, where
\begin{equation}
P_0(x) = e^{x_1}
\label{20}
\end{equation}
is not contained in $\det g^{-1}$ as a factor. This prepotential gives rise to the oscillator 
potential.

Finally we mention that $e^{-\chi}$ is the ground state wave function of the Schr\"odinger 
operator, as follows from (\ref{10}).

The expression (I, 6.17) for the potential $W(x)$ contains a term linear in $\chi$
\begin{equation}
- \sum_{a,b} \frac{\partial}{\partial x_a} \left( g^{-1}_{ab} \frac{\partial \chi}
{\partial x_b} \right) = 
 - \frac12 \sum^r_{i=1} (\gamma_i - \frac12) \sum_a \frac{\partial}{\partial x_a} r^{(i)}_a(x)
\label{21}
\end{equation}
 Each divergence
\begin{equation}
\sum_a \frac{\partial}{\partial x_a} r^{(i)}_a (x) = C^{(i)}
\label{22}
\end{equation}
ought to be a constant to render the operator (\ref{8}) formally self-adjoint after 
subtraction of a constant. From now on we shall dismiss all constant terms in $W(x)$.

We can then write the potential as
\begin{equation}
W(x) = \sum_{i,j} \gamma_{ij} R_{ij}
\label{23}
\end{equation}
\begin{equation}
R_{ij} = \sum_{a,b} g^{-1}_{ab} (x) \frac{\partial \ln P_i}{\partial x_a} \frac
{\partial \ln P_j}{\partial x_b}
\label{24}
\end{equation}
\begin{equation}
\gamma_{ij} = \frac14 (\gamma_i\gamma_j - \frac14) \quad (i,j \not= 0).
\label{25}
\end{equation}
In the cases of this article
\begin{equation}
R_{ij} = {\rm const~if}\, i \not= j
\label{26}
\end{equation}
If we then set
\begin{equation}
\gamma_i = - \nu_i + \frac12 \quad (i \not= 0)
\label{27}
\end{equation}
we obtain
\begin{equation}
W(x) = \sum^r_{i=1} \gamma_{ii}R_{ii}
\label{28}
\end{equation}
with
\begin{equation}
\gamma_{ii} = \frac14 \nu_i (\nu_i-1)
\label{29}
\end{equation}

\section{A model of Calogero type}
In the Schr\"odinger operator for the $A_3$ Calogero model
\begin{equation}
H_{\rm cal} = \frac12 \sum^4_{i=1} \left( - \frac{\partial^2}{\partial x^2_i} + \omega^2x^2_i \right)
+ g_1 \sum_{i<j} (x_i-x_j)^{-2}
\label{30}
\end{equation}
we separate the centre-of-mass part by introducing relative coordiantes
\begin{eqnarray}
X = x_1 + x_2 + x_3 + x_4 \nonumber \\
y_i = x_i - \frac14 X, \quad i \in \{1,2,3 \}
\label{31}
\end{eqnarray}
Following \cite{3} we introduce elementary symmetric polynomials as new coordinates 
instead of the $\{\tau_i \}$
\begin{eqnarray}
\tau_2 &=& \sum_{1 \le i < j \le 4} y_i y_j \nonumber \\
\tau_3 &=& \sum_{1 \le i < j < k \le 4} y_i y_j y_k \nonumber \\
\tau_4 &=& y_1 y_2 y_3 y_4 
\label{32}
\end{eqnarray}
Using the ground state wave function
\begin{equation}
\Psi_0(x) = V(x)^{\nu_1} \exp \left\{ - \frac12 \omega \sum^4_{i=1} x^2_i \right\}
\label{33}
\end{equation}
with $\nu_1$ any root of
\begin{equation}
g_1 = \nu_1(\nu_1-1) 
\label{34}
\end{equation}
and $V(x)$ the Vandermonde determinant,
we gauge $H_{\rm cal}$ to
\begin{eqnarray}
\tilde{H}_{\rm cal} &=& \Psi_0(x)^{-1} H_{\rm cal} \Psi_0(x) \nonumber \\
&=& \tilde{H}_{\rm c.o.m.} + \tilde{H}_{\rm rel} + {\rm const}
\label{35}
\end{eqnarray}
Then (see \cite{3} and I (7.1) - (7.6) with $s=0, \, w_2 = w_3 =w_4 =1$) we get
\begin{eqnarray}
2 \tilde{H}_{\rm rel} &=& - \sum^4_{a,b=2} g^{-1}_{ab} \frac{\partial}{\partial \tau_a} 
\frac{\partial}{\partial \tau_b} \nonumber \\
& & +\mbox{ first order differential operator}
\label{36}
\end{eqnarray}
with
\begin{equation}
g^{-1} = \left( \begin{array}{ccc}
-2\tau_2 & - 3\tau_3 & - 4\tau_4 \\
- 3\tau_3 & - 4\tau_4 + \tau^2_2 & + \frac12 \tau_2 \tau_3 \\
- 4\tau_4 & + \frac12 \tau_2\tau_3 & - 2\tau_2\tau_4 + \frac34 \tau^2_3 
\end{array} \right)
\label{37}
\end{equation}
The first order differential operator is discarded.

The relative motion Hamiltonian (\ref{36}) reflects the symmetry of $H_{\rm cal}$ (\ref{30}) 
under 
\begin{equation}
x_i \to - x_i
\label{38}
\end{equation}
namely it is symmetric under
\begin{equation}
\tau_2 \to \tau_2, \; \tau_3 \to - \tau_3, \; \tau_4 \to \tau_4
\label{39}
\end{equation}
Thus it leaves polynomial spaces that are even or odd in $\tau_3$ separately invariant. 
Now we reduce these spaces and consider polynomials only that are \underline{even} in $\tau_3$:
\begin{equation}
\tau_2 = \lambda_2, \; \tau^2_3 = \lambda_3, \; \tau_4 = \lambda_4 
\label{40}
\end{equation}
\begin{eqnarray}
\frac{\partial}{\partial \tau_3} &=& 2 \lambda_3^{\frac12} \frac{\partial}{\partial \lambda_3} 
\nonumber \\
\frac{\partial^2}{\partial \tau^2_3} &=& 4 \lambda_3 \frac{\partial^2}{\partial \lambda^2_3} + 
2 \frac{\partial}{\partial \lambda_3}
\label{41}
\end{eqnarray}
(this substitution (\ref{40}) was invented in \cite{4,5}). Multiplying the second row and 
column with $2 \lambda_3^{\frac12}$ we obtain
\begin{equation}
g^{-1}_{ab} \to G^{-1}_{ab}
\label{42}
\end{equation}
Whereas the quadratic differential operator (\ref{36}) with $g^{-1}_{ab}$ as in (\ref{37}) and
\begin{eqnarray}
P_1(\tau) &=& - 4 \det g^{-1} \quad (\textrm{I, (7.22), and appendix (A.1)}) \\
\label{43} 
P_0(\tau) &=& e^{\tau_2} \\
\label{44}
r^{(1)} &=& (-12, 0, -2\tau_2) \\
\label{45}
r^{(0)} &=& (-2\tau_2, - 3 \tau_3, - 4 \tau_4)
\label{46}
\end{eqnarray}
allows to reconstruct $H_{\rm cal}$ completely, we have now an additional factor in
\begin{equation}
\det G^{-1} = 4 \lambda_3 (- \frac14) P_1 (\lambda)
\label{47}
\end{equation}
($P_1(\lambda)$ is obtained from $P_1(\tau)$ (A.1) by substitution (\ref{40})).

The polynomial spaces (\ref{1}) are now such that
\begin{equation}
n = 3, \; p_1 = p_3 = 1, \; p_2 = 2
\label{48}
\end{equation}
For the first order differential operators (\ref{2}) we have
\begin{eqnarray}
[\vec{\alpha};a] \in  & & \Bigg\{ [\vec{0};a] \, {\rm all} \; a; \nonumber \\
& & [e^{(a)};1], \, [e^{(a)};3], \; a \in \{1,3\}; \nonumber \\
& & [e^{(a)};2], \, {\rm all} \, a; \nonumber  \\
& & [e^{(a)} + e^{(b)};2], \, a,b \in \{1,3\} \Bigg\}
\label{49}
\end{eqnarray}
If we multiply two of these we obtain possible second order differential operators (\ref{3}), 
but in addition we have 
\begin{equation}
[\vec{\beta};a,b] \in \{[e^{(2)};a,b] \quad a,b \in \{1,3\} \}
\label{50}
\end{equation}

Now the program of the preceding section is set in action: From
\begin{eqnarray}
P_1(\lambda) &=& - \frac{1}{\lambda_3} \det G^{-1} \label{51} \\
P_2(\lambda) &=& \lambda_3 \label{52} \\
P_0(\lambda) &=& e^{\lambda_2} \label{53}
\end{eqnarray}
we obtain
\begin{eqnarray}
r^{(1)} &=& (-12, 0, - 2 \lambda_2) \label{54} \\
r^{(2)} &=& (-6, 4(\lambda_2^2-4\lambda_4), \lambda_2) \label{55} \\
r^{(0)} &=& (-2\lambda_2, -6\lambda_3, -4\lambda_4) \label{56} 
\end{eqnarray}
and
\begin{eqnarray}
R_{11} &=& \frac{16}{P_1} [2\lambda_2^5 + 16\lambda_2^3\lambda_4 - 96 \lambda_2\lambda^2_4
+ 9 \lambda^2_2\lambda_3 + 108\lambda_3\lambda_4 ] \label{57} \\
R_{22} &=& \frac{4}{\lambda_3} [\lambda_2^2-4\lambda_4] \label{58} \\
R_{00} &=& - 2 \lambda_2 \label{59}
\end{eqnarray}
$R_{11}$ leads us back to the Calogero and $R_{00}$ to the oscillator potential. The new potential 
following from $R_{22}$ is obtained as follows. We factorize $\lambda_3$ (eliminating
$y_4 = - y_1 - y_2- y_3$)
\begin{equation}
\lambda_3 = \tau^2_3 = (y_1+y_2)^2 (y_1+y_3)^2 (y_2+y_3)^2
\label{60}
\end{equation}
and perform a fractional decomposition
\begin{equation}
\frac{Q}{\lambda_3} = \sum_{1 \le i < j \le 3} \quad (y_i + y_j)^{-2}
\label{61}
\end{equation}
We find
\begin{equation}
Q = \lambda^2_2 - 4\lambda_4
\label{62}
\end{equation}
so that
\begin{equation}
R_{22} = 16 \sum_{\mbox{\tiny{3 indep. terms}}} (x_i+x_j - x_k-x_l)^{-2}
\label{63}
\end{equation}
Using (\ref{29}) we obtain finally the new model
\begin{eqnarray}
H_{\rm cal} &=& \frac12 \sum^4_{i=1} \left( - \frac{\partial^2}{\partial x_i^2} +
\omega^2x^2_i \right) \nonumber \\
&+& g_1 \sum_{1 \le i < j \le 4} (x_i-x_j)^{-2} \nonumber \\
&+& g_2 \sum_{\mbox{\tiny{3 indep. terms}}} (x_i+x_j - x_k-x_l)^{-2}
\label{64}
\end{eqnarray}
with
\begin{equation}
g_2 = 2\nu_2 (\nu_2-1)
\label{65}
\end{equation}

In the course of this derivation we have reduced the polynomial spaces of the standard $A_3$ 
Calogero model to spaces of polynomials that are even in $\tau_3$. But afterwards we multiply 
them with
\begin{equation}
\lambda_3^{\frac12 \nu_2} = \tau_3^{\nu_2} \quad (\tau_3 > 0)
\label{66}
\end{equation}
So for $\nu_2=0$ we have recovered the even and for $\nu_2=1$ the odd spaces. In these cases 
$g_2=0$ as shown by (\ref{65}). Thus the new potential arises by an interpolation between 
even and odd spaces.

\section{A model of Sutherland type}
Having constructed a model of Calogero type the existence of a corresponding model of 
Sutherland type can be expected. We start from our formulas for the $N=4$ Sutherland model 
in (I, Section 7) and put everywhere
\begin{equation}
w_i=1, \quad i \in \{2,3,4\}, \quad s = 2
\label{67}
\end{equation}
Again we change variables as in (\ref{40}), (\ref{41}) and obtain
\begin{equation}
\det (G^{-1}) = 4 \lambda_3 (- \frac14) P_1 (\lambda)
\label{68}
\end{equation}
as in (\ref{47}), and
\begin{equation}
P_2(\lambda) = \lambda_3
\label{69}
\end{equation}
as in (\ref{52}). $P_1(\lambda)$ is given as in the appendix (A.2). There is no place for a $P_0(\lambda)$. The reader is advised to try out 
what happens with a $P_0(\lambda)$.

Calculations become very lengthy. We find 
\begin{eqnarray}
r_a^{(1)} &=& (2(a_7-1)(a_7-3)\lambda_4 - 2(a_7+5)\lambda_2-12, \nonumber \\
& & - 24 \lambda_3, \nonumber \\
& & + 2(a_7-9) \lambda_4-2\lambda_2)
\label{70}
\end{eqnarray}
\begin{eqnarray}
r_a^{(2)} &=& (-(a_7-1)(a_7-3)\lambda_4 + (a_7-7)\lambda_2-8, \nonumber \\
& & +4 [(a_7-1)\lambda_4 - \lambda_2]^2 - 8\lambda_3 - 16\lambda_4, \nonumber \\
& & - (a_7+3) \lambda_4+\lambda_2)
\label{71}
\end{eqnarray}
The parameter $a_7$ appears only in the Lie algebraic formulation of the same ($a_7$ independent) 
Sutherland model.

The potentials $R_{ij}$ come out as
\begin{eqnarray}
R_{11} & = & \frac{16}{P_1} \{R_1\} \\ \label{72}
& &\textrm{with $R_1$ given in the appendix (A.3)} \nonumber \\
R_{22} &=& \frac{4}{\lambda_3} \{ (a_7-1) \lambda_4 - \lambda_2]^2 - 4\lambda_4 \} \nonumber \\
& & + \mbox{ irrelevant terms}
\label{73}
\end{eqnarray}
and $R_{12}$ is constant (irrelevant). We introduce (see (I), eqs. (7.16)-(7.18)) the shorthand 
for the trigonometric $S_4$--symmetric functions
\begin{eqnarray}
\sigma_2 &=& \sum \sin y_i \sin y_j \cos y_k \cos y_l \nonumber \\
&=& - \frac12 \sum_{1\le i< j\le 3} \sin^2(y_i+y_j)
\label{74}
\end{eqnarray}
\begin{eqnarray}
\sigma_3 &=& \sum \sin y_i \sin y_j \sin y_k \cos y_l \nonumber \\
&=& - \prod_{1\le i< j\le 3} \sin (y_i+y_j)
\label{75}
\end{eqnarray}
\begin{equation}
\sigma_4 = \sin y_1 \sin y_2 \sin y_3 \sin y_4 
\label{76}
\end{equation}
Since
\begin{equation}
\lambda_3 = \sigma^2_3
\label{77}
\end{equation}
we study the fractional decomposition
\begin{equation}
\frac{Q}{\lambda_3} = \sum_{1 \le i < j \le 3} (\sin (y_i+y_j))^{-2}
\label{78}
\end{equation}
For $Q$ we find
\begin{equation}
Q = (\sigma_2-2\sigma_4)^2 - 4\sigma_4 + \sigma^2_3
\label{79}
\end{equation}
On the other hand we get by insertion of $\lambda_2, \lambda_4$ ($\xi = 2\lambda_2, 
\eta_4 = 4\lambda_4$ in I (7.9) and (7.11))
\begin{equation}
R_{22} = \frac{4Q}{\lambda_3} + {\rm const}
\label{80}
\end{equation}
Thus we end up with a Sutherland model
\begin{eqnarray}
H_{\rm suth} &=& \frac12 \sum^4_{i=1} \left( - \frac{\partial^2}{\partial x^2_i} +
\omega^2 x^2_i \right) \nonumber \\
& & + g_1 \sum_{1 \le i < j \le 4} (\sin (x_i-x_j))^{-2} \nonumber \\
& & + \frac14 g_2 \sum_{{\textrm{\tiny{3 cases}}}} (\sin \frac12 (x_i + x_j - x_k - x_l))^{-2} 
\label{81}
\end{eqnarray}
with $g_2$ as in (\ref{65}).
The remarks made at the end of the preceding section remain true in this case.

\section{Concluding remarks}
Our technique can also be applied to the $A_2$ model whereas the cases $A_{N-1}$, $N>4$, 
cause hitherto unsolved problems. In the latter cases the form of the trigonometric models 
in the variables $\{\sigma_n\}^N_2$ analogous to (\ref{74}) -- (\ref{76}) is also known \cite{8}. The 
trigonometric variables $\{\eta_n\}^{N-1}_1$ introduced in \cite {3} in which all $A_{N-1}$ 
models were expressed, are algebraically related with the $\{\sigma_n\}^N_2$ 
but the explicit form of this relation is not known. The trigonometric variables introduced 
in \cite{4} eqns. (2.17), (2.18) are $\Re{\eta_1} , \Im{\eta_1}$ respectively, 
those of \cite{4}, eqns. (3.4), (3.5) are still different.

The rational $A_2$ and $G_2$ models are equivalent if
\begin{equation}
\nu(A_2) = \nu(G_2) + \mu(G_2)
\label{82}
\end{equation}
 where the respective coupling constants are (see \cite{4})
\begin{eqnarray}
g(A_2) &=& \nu(A_2) (\nu(A_2)-1) \label{83} \\
g(G_2) &=& \nu(G_2) (\nu(G_2)-1) \label{84} \\
g_1(G_2) &=& 3 \mu(G_2) (\mu(G)-1) \label{85} 
\end{eqnarray}
Indeed, their Lie algebraic version (\cite{4}), eqns. (2.8), (4.5)) are identical if 
(\ref{82}) is satisfied. The trigonometric versions are, however, inequivalent.

Applying our technique to the $A_2$ model gives the $G_2$ model, both in the rational and 
the trigonometric cases. Since the $A_2$ model is obtained from the $G_2$ model by 
specialization of one coupling constant, we should denote both models together as $AG_2$. 
The extension of the $A_3$ model found by us could to be denoted $AG_3$ correspondingly. 
The Lie algebras $A_3$ and $D_3$ are identical. But $A_3$ and $D_4$ models involve four 
particles, the center-of-momentum motion of the $D_4$ model is not separable. Its potential 
is \cite{1}
\begin{eqnarray}
g\sum_{1\leq i < j \leq 4} [v(x_i-x_j)+v(x_i+x_j)] \label{86}
\end{eqnarray}
whereas our $AG_3$ model has potential 
\begin{eqnarray}
\sum_{1\leq i < j \leq 4} [g_1v(x_i-x_j) + \frac14 g_2 v(y_i+y_j)] \label{87}
\end{eqnarray}
They are obviously related, but (\ref{87}) differs from (\ref{86}) by independence of 
the coupling constants and translational invariance.

Finally we want to mention that we have derived the $AG_3$ model also by the use of the 
coordinates $\mu_1,\mu_2,\mu_3$ from
\begin{eqnarray}
\mu_1 & = & \sum_{i=1}^{4} y_i^2 = -2\tau_2 \label{88} \\
\mu_2 & = & \sum_{1\leq i < j \leq 4} y_i^2 y_j^2 = +2\tau_4+\tau_2^2 \label{89} \\
\mu_3 & = & \sum_{1\leq i < j < k \leq 4} y_i^2 y_j^2 y_k^2 = -2\tau_2\tau_4+\tau_3^2 \label{90} \\
\mu_4 & = & y_1^2 y_2^2 y_3^2 y_4^2 = \tau_4^2 \label{91} 
\end{eqnarray}
so that 
\begin{equation}
\mu_4 = \frac14 (\mu_2 - \frac14 \mu_1^2)^2 \label{92} 
\end{equation}
This derivation goes along the same lines but is more laborious. It is remarkable that 
in this case one has to use an infinite ensemble of flags of polynomial spaces
\begin{equation}
V_N=\textrm{span}\{\mu_1^{n_1}\mu_2^{n_2}\mu_3^{n_3} ; \quad n_1 p_1 + n_2 p_2 + n_3 p_3 \leq N \} \label{93} 
\end{equation}
namely one flag for each admissible triplet $(p_1,p_2,p_3) \in \bbbn^3$. In the following table we give some of these triplets.
The minimal admissible triplet is $(2,3,5)$.
\begin{center}
%
\begin{tabular}{|c|c|c|} \hline
(1,2,4) & (2,3,5) & (3,5,10) \\ \hline
(1,3,6) & (2,3,6) & (3,7,14) \\ \hline
(1,4,8) & (2,4,5) & (3,8,16) \\ \hline 
(1,5,10) & (2,5,6) & (3,10,20) \\ \hline
(1,6,12) & (2,5,10) & (3,11,22) \\ \hline
\vdots & (2,6,7) & \vdots \\ \hline
(1,$p_2$,$2p_2$) & \vdots  & (3,$p_2$,$2p_2$) \\ 
\end{tabular} \\
{Table 1: Table of admissible triplets $(p_1,p_2,p_3)$.}
\end{center} 
Criterion for admissibility is that the differential operator $D$ (7),(8) maps $V_N$ in 
$V_N$ and possesses nontrivial eigenvectors.
\section{Note added}

Recently we were able to prove that the translation invariant four--particle 
model $AG_3$ considered here is equivalent to the translation non-invariant 
three--particle model belong to $B_3$. The arguments rely on a discussion of 
the underlying Coxeter groups and their invariants.  

\section{Appendix: Some formulas}
In the case of the Calogero $A_3$ model we have from (I,(7.22))
\begin{eqnarray}
P_1(\tau) = 27\tau_3^4-256\tau_4^3+128\tau_2^2\tau_4^2 -16\tau_2^4\tau_4+4\tau_2^3\tau_3^2-144\tau_2\tau_3^2\tau_4 \qquad \textrm{(A.1)} \nonumber
\end{eqnarray}
In the case of the Sutherland $A_3$ model we have (we give these functions for $a_7=3$ only)  
\begin{eqnarray}
P_1(\lambda)= & &- 64\lambda_3\lambda_2^3\lambda_4 - 16\lambda_2^5\lambda_4 - 48\lambda_2
\lambda_4\lambda_3^2 - 144\lambda_2
\lambda_4\lambda_3 + 528\lambda_3\lambda_2
\lambda_4^2 + 256\lambda_3\lambda_2^2
\lambda_4^2 \nonumber \\
 & & - 384\lambda_3\lambda_2\lambda_4
^3 + 128\lambda_2^2\lambda_4^2 - 768
\lambda_2\lambda_4^3 + 27\lambda_3^2 + 192
\lambda_3\lambda_4^2 - 48\lambda_4
\lambda_3^2 + 8\lambda_2^2\lambda_3^2 \nonumber \\
 & & + 48\lambda_3^2\lambda_4^2 - 384
\lambda_3\lambda_4^3 + 192\lambda_3
\lambda_4^4 - 256\lambda_4^3 + 768\lambda_4
^4 + 4\lambda_3^3 - 208\lambda_2^2\lambda
_3\lambda_4 \qquad \textrm{(A.2)} \nonumber \\
 & & + 36\lambda_2\lambda_3^2 - 768
\lambda_4^5 + 256\lambda _4^6 - 16\lambda_2
^4\lambda_4 + 256\lambda_2^3\lambda_4^2 - 1024\lambda_2^2\lambda_4^3 + 144\lambda_2^4\lambda_4^2 \nonumber \\
 & & - 512\lambda_2^3\lambda_4^3+896
\lambda_2^2\lambda_4^4- 768\lambda_2
\lambda_4^5 + 1536\lambda_2\lambda_4^4 + 4\lambda_3\lambda_2^3 + 4\lambda_3\lambda_2^4 \nonumber
\end{eqnarray}
and 
\begin{eqnarray}
R_1=& & 272\lambda_3\lambda_2^3\lambda_4
 + 108\lambda_4\lambda_3 + 56\lambda_2^5\lambda_4 + 216\lambda_2\lambda_4\lambda_
3^2 + 9\lambda_2^2\lambda_3 + 732\lambda_2\lambda_4\lambda_3 + 2\lambda_2^5 \nonumber \\
 & & - 2352\lambda_3\lambda_2\lambda_4
^2 - 1136\lambda_3\lambda_2^2\lambda_4^2 + 1728\lambda_3\lambda_2\lambda_4^3
 - 96\lambda_2\lambda_4^2 + 16\lambda_2^3
\lambda_4 - 736\lambda_2^2\lambda_4^2
\nonumber \\
 & & + 3648\lambda_2\lambda_4^3 - 108
\lambda_3^2 - 972\lambda_3\lambda_4^2 + 216
\lambda_4\lambda_3^2 - 34\lambda_2^2
\lambda_3^2 - 216\lambda_3^2\lambda_4^2 \qquad \qquad \quad \textrm{(A.3)} \nonumber \\
 & & + 1728\lambda_3\lambda_4^3- 864
\lambda_3\lambda_4^4 + 1152\lambda_4^3 - 
3456\lambda_4^4 - 18\lambda_3^3 + 2\lambda
_2^6 + 896\lambda_2^2\lambda_3\lambda_4 \nonumber \\
 & &  - 150\lambda_2\lambda_3^2 + 3456
\lambda_4^5 - 1152\lambda_4^6 + 72\lambda_2
^4\lambda_4 - 1200\lambda_2^3\lambda_4^
2 + 4736\lambda_2^2\lambda_4^3 \nonumber \\
 & & - 600\lambda_2^4\lambda_4^2 + 2240
\lambda_2^3\lambda_4^3 - 4000\lambda_2^
2\lambda_4^4 + 3456\lambda_2\lambda_4^5
 - 7008\lambda_2\lambda_4^4 - 4\lambda_3
\lambda_2^3 \nonumber \\
 & & - 14\lambda_3\lambda_2^4 \nonumber
\end{eqnarray}

\end{document}